\documentclass[useAMS,usenatbib]{mn2e} % mnras paper
\usepackage[varg]{txfonts}
\usepackage{natbib}
\usepackage{graphicx}
\usepackage{txfonts}
\title[Generation of inclined protoplanetary discs and misaligned planets through mass accretion]{Generation of inclined protoplanetary discs and misaligned planets through mass accretion I: Coplanar secondary discs}
\author[M. Xiang-Gruess \& P. Kroupa]{M. Xiang-Gruess$^{1}$ \thanks{E-mail:
mxiang@astro.uni-bonn.de } , P. Kroupa$^{1,2}$ \\
$^{1}$  Helmholtz-Institut f\"ur Strahlen- und Kernphysik, Nussallee 14-16, 53115 Bonn, Germany \\
$^{2}$ Charles University in Prague, Faculty of Mathematics and Physics, Astronomical Institute, V  Hole\v{s}ovi\v{c}k\'ach 2, CZ-180 00 Praha 8, Czech Republic
}

\begin{document}

\date{Accepted . Received ; }

\pagerange{\pageref{firstpage}--\pageref{lastpage}} \pubyear{2015}
\maketitle

\label{firstpage}

\begin{abstract}

We study the three-dimensional evolution of a viscous protoplanetary disc which accretes gas material from a second protoplanetary disc during a close encounter in an embedded star cluster. 
The aim is to investigate the capability of the mass accretion scenario to generate strongly inclined gaseous discs which could later form misaligned planets.
We use smoothed particle hydrodynamics to study mass transfer and  disc inclination for passing stars and circumstellar discs with different masses. We explore different orbital configurations to find the parameter space which allows significant disc inclination generation.

\citet{Thi2011} suggested that significant disc inclination and disc or planetary system shrinkage can generally be produced by the accretion of external gas material with a different angular momentum. We found that this condition can be fullfilled for a large range of gas mass and angular momentum.

For all encounters, mass accretion from the secondary disc increases with decreasing mass of the secondary proto-star. Thus, higher disc inclinations can be attained for lower secondary stellar masses.
Variations of the secondary disc's orientation relative to the orbital plane can alter the disc evolution significantly. 

The results taken together show that mass accretion can change the three-dimensional disc orientation   significantly resulting in strongly inclined discs. In combination with the gravitational interaction between the two star-disc systems, this scenario is relevant for explaining the formation of highly inclined discs which could later form misaligned planets. 

\end{abstract}

\begin{keywords}
planetary systems: formation -- planetary systems: protoplanetary discs 
\end{keywords}

\section{Introduction}

Theories of planet formation have originally been based on our knowledge of the Solar system where the orbits of all planets are nearly coplanar with the Sun's equatorial plane. In the framework of the first theories it was also expected that giant planet formation can only occur beyond the snow line, where sufficient icy material is available to form multi-Earth mass protoplanetary cores which accrete gas rapidly at a later evolutionary stage.   
Two decades after the discovery of the first extrasolar planet 51 Peg b \citep[][]{May1995}, we have an impressive database of almost 3000 confirmed exoplanets with all kinds of interesting characteristics today. For about $10\%$ of all confirmed exoplanets, an observed measure of the angle between their orbital plane and the equatorial plane of their host star \citep[e.g.][]{Win2007, Tri2010} is available. In around $40\%$ of these (single as well as multi-) planetary systems, the angular momentum vector is significantly misaligned with the angular velocity vector of the central star \citep[e.g.][]{Tri2010, Alb2012, Hub2013} which is known as spin-orbit misalignment. 
Due to the observational bias for preferentially detecting massive planets on close orbits about their central star, this fraction of misaligned systems is an upper limit on the fraction of such systems amoungst all planetary systems.
Two different scenarios have been proposed and studied in the past to explain the possible formation and evolution of the so-called misaligned Hot Jupiters. 
As the stellar and disc angular velocities are expected to be aligned in standard planet formation scenarios \citep[e.g.][]{Miz1980, Pol1996, May2002}, the {\it first scenario} assumes an inclination of the planetary orbit relative to the nascent protoplanetary disc.
A possible mechanism to produce high orbital inclinations with respect to the stellar equator  have been presented by e.g. \cite{Tho2003}, who  have studied the evolution of two giant planets in resonance by adopting an approximate analytical expression for the influence of a gas disc in producing orbital migration. Resonant inclinations up to $\approx  60\degr$ could be excited when the eccentricity of the inner planet reaches a threshold value  $\approx   0.6$. Other mechanisms of inclination generation involve a distant companion which can trigger Lidov-Kozai cycles \citep[e.g.][]{Fab2007, Wu2007} or planet-planet scattering and chaotic interactions \citep[e.g.][] {Weid1996,Rasio1996,Pap2001,Nag2008}. When the planet attains a small enough pericentre distance, tidal interaction with the central star will lead to orbital circularization, leaving the planet on a short-period inclined  circular orbit.

However in  a recent study,  \citet{Daw2013} find that large eccentricities,  that are expected to  be associated with such interactions for orbits too wide to be affected by stellar tides,   are seen  mostly in metal rich systems. Accordingly they conclude  that  gentle  disc migration and planet-planet scattering must  both operate during the early evolution of giant planets. Hence, only systems packed with giant planets, which most easily form around metal rich stars, can produce large eccentricities through scattering. They also note a lack of a correlation between spin-orbit misalignment and metallicity. A correlation would however have indicated the operation of such a mechanism. % that can cause misalignment of the disc while simultaneously  allowing disc migration of giant planets into close orbits.

Adjacent to the possible formation of misaligned planetary orbits, the influence of a coplanar protoplanetary disc on an inclined planetary orbit has been explored by several recent studies \citep[e.g.][]{Cre2007, Mar2009, Bit2011, Rei2012, Xia2013}. In almost all cases, simulations showed damping for both eccentricity and inclination with the planets circularising  in the disc after a few hundred orbits. Only in the case of very high initial inclinations $>70^{\degr}$ and planet masses of one Jupiter mass, the timescale of realignment with the gas disc was found to be  comparable with the disc lifetime \citep[][]{Xia2013}. These results taken together imply that the first scenario involving a coplanar disc and an inclined planetary orbit fails to explain the existence of misaligned planets in most cases. 

The {\it second scenario} for the formation of misaligned planets, that is consistent with inward migration driven by a disc, may be connected with the possibility that the protoplanetary disc itself is inclined. This could happen either through gravitational perturbation by a stellar companion or stellar flyby, but also through accretion of gas material with a significantly different angular moment vector. Thereby, the pertubation event may occur either before or after  planet formation  \citep[e.g.][]{Bat2010,Thi2011, Xia2016}.

In this regard we note that in general, protostars are not  formed in isolation, but are  bound in stellar clusters where gravitational interactions  are common \citep[][]{Lad2003, Meg2016}. Observations of stellar clusters have provided a global initial mass function (IMF) which describes the distribution of the stellar masses \citep[e.g.][]{Kro2002, Cha2003, Bas2010, Kro2013}. Taking into account the mass distribution in a stellar cluster, it can be very common that close encounters occur for different mass stars \citep[][]{Thi2005}, especially during the compact pre-gas expulsion phase \citep[][]{Mar2012}. 
The effects of single close encounters on circumstellar discs can be very strong as the perturbing star can destroy the global structure and dynamics of a disc. These gravitational effects of a close stellar encounter on a circumstellar disc have been studied analytically as well as numerically by many groups in the past years \citep[e.g.][]{Cla1993, Kor1995, Kob2001, Pfa2003, Fra2009}. For different perturber orbits, these studies have mainly concentrated on equal-mass stars, disc mass loss or disc sizes. 
More recent studies of protoplanetary discs and planetary systems affected by single stellar flybys have investigated a parameter range spanning the entire IMF \citep[e.g.][]{Mal2011, Bre2014}. 
Several groups have performed N-body simulations of stellar clusters \citep[e.g.][]{Spu2009, Mal2007, Mal2011, Olc2012, Cra2013, Vin2015} in order to study rates and properties of close encounters and their influences on disc sizes and mass loss,  but also on planetary systems \citep[e.g.][]{Li2015, Por2015, Par2016}. 

In this context, gas accretion during close encounters in stellar clusters have been studied in a simplified way by \citet{Pfa2008a, Pfa2008b}.
Stellar encounters can also play a significant role for massive gaseous discs as they can lead to gravitational instabilities and subsequent fragmentation of the discs.  
In the past few years, observations were able to detect signatures of possible accretion processes around young stellar objects \citep[e.g.][]{Liu2016} which could be explained by gravitational fragmention of the protoplanetary discs.

Recently, \citet{Xia2016} performed a parameter study of single stellar flybys in order to study the evolution of inclination in hydrodynamical discs and succeeded in generating disc inclinations up to $60\degr$ with retrograde inclined orbits. 

The evolution of protoplanetary discs which undergo periods of mass accretion has been studied by only a few groups so far. \citet{Thi2011}  showed that the capture of gas from accretion envelopes by  a protoplanetary disc could cause it to become significantly misaligned with respect to its original plane. In this scenario, the original disc and/or already existing young planetary system shrinks possibly explaining the misaligned hot (close-in) Jupiters. It is of interest to remark that  misaligned gas discs can easily be produced  when  interactions  with the environment out of which they form are considered. However, this result has been questioned recently by \citet{Pic2014} who studied the 3D evolution of circumstellar discs during stellar flybys including mass transfer between the two gas discs around the stars. In their simulations, disc inclinations of $<10\degr$ only were generated. 
The mass accretion process has also been studied for the exploration of other relating topics. \citet{Wij2016} have investigated whether possible face-on accretion onto protoplanetary discs can produce so-called 'second generation' globular cluster stars \citep[e.g.][]{Pio2007, Bas2013}. \cite{Hen2016} have studied the formation of spiral arms in protoplanetary discs which accrete gas material at their outer disc boundaries. 

So far, the only studies of the mass accretion scenario which investigate the possible generation of inclined gaseous discs are those by \cite{Thi2011} and \cite{Pic2014}. 
 In view of the above discussion about misaligned planets and the prevailing controversial predictions made by the two groups with respect to the mass accretion scenario, it is important to investigate the evolution of a protoplanetary disc undergoing mass accretion during a close encounter with another star in more detail as previous studies \citep[e.g.][]{Xia2014} have shown that a gaseous circumstellar disc is able to control the evolution of the whole disc-planet system. 
 
In this paper, we study the three-dimensional evolution of a viscous protoplanetary disc which is perturbed by a passing star and its surrounding gas disc. The aim is to answer the question whether single stellar flybys in combination with mass transfer between two protoplanetary discs is capable to generate the full range of disc inclinations and hence observed misalignment angles of exoplanets. 
We make the simplification of neglecting the disc self-gravity which results in a significant reduction in computational resource requirements. This is mainly motivated by the fact that disc masses are usually very small compared to the central stellar mass \citep[e.g.][]{And2013}.

In studying the disc inclination, we note that warping of protoplanetary discs can generally occur as a result of tidal interaction with a binary star or if the disc accretes matter with angular momentum misaligned with that of the star \citep[e.g.][]{Ter1993, Pap1995, Bat2010, Thi2011}. Possible warping of the disc in the presence of a binary star star has been studied by e.g. \cite{Xia2014, Lub2016}.
So far, simulations of single stellar flyby events in the absence of mass accretion processes \citep[][]{Xia2016} did not show any indications for disc warping. In the mass accretion scenario, 
the extent of warping for different orbital parameters has still not been studied in detail. However the potential warping of a protoplanetary disc is relevant in the discussion of planets since it has direct consequences on the formation and three-dimensional evolution of planets in these systems \citep[see also][]{Ter2013}.

It has been shown that smoothed particle hydrodynamics (SPH) simulations are capable of being applied to the problem of three-dimensional evolutions of gaseous discs rather than grid-based simulation methods. In contrast to grid-based methods, SPH simulations adopt a Lagrangian approach and can therefore be readily used to simulate a gas disc  with a free boundary   that undergoes a large amount of movement  in three dimensions with the disc having the freedom to change its shape. As it is suitable for the system that we aim to study, we adopt this approach here. 
In Section \ref{sec:sim_details}, we describe our simulation technique, giving details of the hydrodynamical treatment of the protoplanetary discs. The general setup of the simulations is discussed in Section \ref{sec:IC}. In addition, we discuss the properties of realistic close encounters in stellar clusters in more detail in Section \ref{sec:orbit}.
We go on to present the simulation results in  Section \ref{sec:results}.
Finally, we summarize and discuss our results in Section \ref{sec:conclusions}.

\section{Simulation details}\label{sec:sim_details}

We  have applied a modified version of the publicly available code {\rm GADGET}-2  \citep[][]{Spr2005} for the simulations shown in this paper. The advantage of GADGET-2 is its capability to simulate fluids as well as distinct massive bodies. In our scenario, the two encountering massive stars are computed by N-body particles while their surrounding gaseous discs are modelled by SPH particles.  
As we are interested in the evolution of the primary disc and aim to perform fast analyzing and visualizing steps, we chose a Cartesian coordinate system with its origin coinciding with the centre of mass of the primary star and the $(x,y)$ plane characterizing the initial midplane of the primary disc. This choice can be done without loss of generality if the well-known indirect term \citep[][]{Pap1995} which takes into account the acceleration of the coordinate system is included in the equation of motion. 
For our chosen coordinate system, the primary star of mass $M_{s1}$ is kept fixed throughout the simulation and the secondary star (stellar perturber) of mass $M_{s2}$ is allowed to move as a massive body. 
%We adopt spherical polar coordinates $(r, \theta,\phi)$ with origin at the centre of mass of the central star.  The associated Cartesian coordinates $(x,y,z)$ are such that the $(x,y)$ plane coincides with the initial midplane of the primary disc around the central star.

The total unsoftened gravitational potential $\Psi$ at a position ${\bf  r}$ is given by \citep[see also][]{Pap1995, Lar1996, Xia2016}
\begin{eqnarray}
 \Psi({\bf  r})&=& -\frac{G M_{s1} }{|{\bf  r}|} - \frac{G M_{s2} }{|{\bf  r}-{\bf  D}|} + \frac{G M_{s2} {\bf  r}\cdot {\bf D} }{|{\bf  D}|^3} \ .
\label{eq:Pot}\end{eqnarray}
Contributions from the primary star and secondary star, with  position  ${{\bf  D}}$, are included. The last term in Equation (\ref{eq:Pot}) accounts for the acceleration of the coordinate system as the origin of the coordinate system  moves with the primary star.

As in all numerical simulations involving collisionless dynamics, the gravitational potential has to be softened for small interaction distances in order to avoid infinite small time steps. This is realised in GADGET-2 by employing the so-called spline kernel \citep[see also Eq. (4) of ][]{Spr2005}. 
The only gravitational softening that is included in our simulation is for the gravitational interaction between the SPH particles and the two stellar particles.
%Thus $G M_*/|{\bf r}|$ is replaced by $G M_*/\sqrt{|{\bf r}|^2+ \varepsilon_B^2}$ in equation (\ref{Pot}).
%In addition  $|{\bf r}-{{\bf r}}_{p,i}|$ and $|{\bf r}-{\bf D}|$
%are replaced by $\sqrt{(|{\bf r}-{{\bf r}}_{p,i}|^2+ \varepsilon_{p}^2 )}$ and $\sqrt{( |{\bf r}-{\bf D}|^2+  \varepsilon_{p}%^2 )}$ respectively.
We applied fixed  softening lengths $\varepsilon_{s1}=\varepsilon_{s2}=0.1$ internal length units for both stars.
As self-gravity is expected to play a minor role in low-mass discs, it is neglected in the simulations. 

 The numerical simulations include a gas accretion routine onto the two stellar particles.  In this respect, a gas particle has to approach a star within a so-called accretion radius and fullfill a series of conditions before it can be accreted by the star. These conditions are crucial since they make sure that the gas particle is indeed bound to the star. In addition, they also test whether the specific angular momentum of the gas particle about the star is sufficiently small in order to prevent the gas particle from forming a circumstellar orbit.
Details of the numerical implementation of the accretion process can be found in \citet{Bat1995}. 
The outer accretion radius of the stars was fixed during the simulation to be $R_a=0.05$ internal length units. The inner accretion radius was taken to be 0.5 of the outer accretion radius. 

The hydrodynamical simulation of the two gaseous discs was performed by adopting a locally isothermal equation of state (EOS) which is characterized by the disc aspect ratio $h=H/r=0.1$. $H$ is the circumstellar disc scale height and $r=|{\bf  r}|$ the distance to the host star. The soundspeed in the disc is given by $c_s=h |{\bf v}_\varphi|$, with ${\bf  v}_\varphi$ being the rotational velocity vector in the circumstellar disc. The pressure is given by  $p=\rho c_s^2$. Thus, the temperature in the disc is $\propto r^{-1}$.
For the SPH calculations, the number of nearest neighbours was set to $40 \pm 5$. 
The artificial viscosity parameter $\alpha$ of GADGET-2 \citep[see equations (9) and (14) of][]{Spr2005} was taken to be $\alpha =0.5$. A detailed discussion of our applied $\alpha$ parameter can be found in \citet{Spr2005} and \cite{Xia2016}. 
To reduce artificially induced  angular momentum transport in the presence of shear flows, a viscosity-limiter has been applied which is especially important for the study of Keplerian discs.
In addition, we include a routine to allow gas particles to switch their host star. In these cases, the hydrodynamical parameters are computed by employing the coordinates and mass of the new host star. 
%When studying stellar encounters involving gas envelopes or discs, the possibility of a gas particle becoming bound to the stellar perturber has to be included. In our case, the gas particles are assigned to the central star at start of the simulations. We include an additional routine to allow gas particles to be assigned to the stellar perturber if their distance $r$ to the central star is larger than 50 au and if they are closer to the perturber than to the central star. In this case, $r$ is replaced by its distance to the perturber when the hydrodynamical parameters are computed.
We found that the application of this procedure did not affect the global outcomes of the simulations.

The gravitational effect of the discs on the stars is neglected. The internal unit of length is $5\ \mathrm{au}$ and the internal time unit is the orbital period at $a=5\ \mathrm{au}$ which is 11.2 yr.  The dimensionless units and their corresponding physical quantities are listed in Table \ref{tab:units} and adopted  for all plots shown in this paper.
 
\begin{table}
 \begin{center}
\begin{tabular}{|c|c|}
\hline
Internal unit & Physical unit  \\
\hline
length & $5\ \mathrm{au}$ \\
time & orbital period ($5\ \mathrm{au}$) = $11.2\ \rmn{yr}$    \\
%20 & 0.89 & 200 \\
\hline
\end{tabular}
\end{center}
\caption{Dimensionless and corresponding physical units applied in all simulations.}
\label{tab:units}
\end{table}

\section[]{Initial conditions} \label{sec:IC}

We study a system composed of two massive stars both surrounded by a Keplerian gaseous disc. The primary star, of mass $M_{s1}=1\ M_{\sun}$, is fixed in the centre of origin throughout the simulation. The primary disc mass is $0.01\ \rmn{M_{\sun}}$ and the primary disc radius is $40\ \rmn{au}$. 
 The comparably small radius in contrast to observed discs with radii of up to a few 100 au is due to numerical resolutional reasons. For a given disc mass and total number of SPH particles, the smoothing lengh at any position in the disc is proportional to $\rho^{-{\frac{1}{3}}}$, with $ \rho$ being the local mass density. With increasing distance to the central star, the local mass density decreases and the smoothing length increases accordingly. In case of a large disc radius, the smoothing length would become extremely large making the hydrodynamical calculations unphysical. 

At simulation start, the midplane of the primary disc is in the $(x,y)$ plane. 
The secondary star follows a parabolic orbit about the primary star. In this paper, we mainly concentrate on the configuration with the initial secondary disc being in the orbital plane of the secondary star.

Based on the numerous surveys of stellar clusters \citep[e.g.][]{Thi2005, Olc2012, Pfa2013, Vin2015}, it is now widely accepted that a star can undergo multiple encounters with other stars which can lead to minor changes of their surrounding discs up to complete disc destruction. The fate of the circumstellar discs is dependent on the stellar mass ratios, the distance between the stars during the close encounter and the specific orbital parameters of the encounter.   
In \citet{Xia2016}, the probability of solar-like stars to undergo a close encounter at a distance of 100-1000 au with another star with at least 5 $\rmn{M_{\sun}}$ has been determined to be roughly 1/4. 

 But note that in stellar clusters of lower masses and densities, stars with at least 5  $\rmn{M_{\sun}}$ are not common \citep[][]{Weid2013}. However, the here studied mass transfer is a general phanomenon which appears for lower stellar masses too. The results are mainly dependent on the stellar and disc mass ratios and not on the actual absolut masses. Thus, in a lower density region, a close encounter between a 0.1 and 0.5 $\rmn{M_{\sun}}$ star with their circumstellar gaseous discs would also lead to mass transfer and the resulting changes of the discs. The applicability of our studied gas accretion scenario will be discussed in detail in the conclusion.

Probabilities of encounters at distances $>100$ au larger than 20\% have also been found for a variety of star clusters by \citet{Thi2005}. These relatively high probabilities show that close encounters with stellar mass ratios significantly different to 1 have to be studied, too. 
Similarly to \citet{Xia2016}, we will perform simulations with perturber masses in the range [1:10] $\rmn{M}_{\sun}$.

The surface mass density of both discs was chosen to be proportional to $R^{-1/2}$, where $R$ is the radial coordinate of a point in the midplane. The disc radii are given by $R_{d1}$ and $R_{d2}$, respectively.
%\begin{eqnarray}
%\Sigma=\Sigma_0 R^{-1/2}. \label{eq:sigma}
%\end{eqnarray}
%Here $\Sigma_0$ is a constant and $R$ is the radial coordinate of a point in the midplane.
%This applies to  the radial domain $[0, R_d]$, where $R_{d}=R_{out}$ is the disc's radius. The disc mass is given by
%\begin{eqnarray}
%M_d=2\pi \int_{R_{\rmn{in}}}^{R_{\rmn{out}}} \Sigma(r)r dr=\frac{4}{3} \pi \Sigma_0 R_{\rmn{out}}^{3/2}\ ,
%\end{eqnarray}
%which is used to determine $\Sigma_0.$
The disc mass is a fundamental property which plays a key role in the context of planet formation and evolution. In theoretical studies, it has been common to assume a fundamental scaling relationship between the disc mass and the stellar mass \citep[e.g.][]{Lau2004, Kor2006}. The observational measurement of the gas disc mass is challenging in both the early protoplanetary phase \citep[e.g.][]{Kam2011, Mio2014} and the late debris phase \citep[e.g.][]{Pas2006}. As the disc mass evolves with time, the disc-stellar mass ratio is not constant, but it depends on the age/phase of the protoplanetary disc. It is theoretically often assumed that the disc mass scales with the stellar mass as $M_d \propto M_*^{\alpha_D}$, where $\alpha_D=[0:2]$ \citep[e.g.][]{Ida2005,Ali2011}. Adjacent to the disc mass evolution, observational surveys have also confirmed the relationship between disc and stellar mass \citep[e.g.][]{Sch2006, Sch2009, And2013}. While \citet{Ali2011} determined $\alpha_D \approx  1.2$ in accordance to the observational data, more recent observations \citep[e.g.][]{Pas2016} found $\alpha_D=1.3-1.9$ for young ($\approx 2$ Myr) protoplanetary discs. Taking into account the newest results for young gas discs, we have performed simulations with disc masses 
\begin{equation}
 M_d = 0.01 M_*^{1.5}\ , \label{eq:m_d}
\end{equation}
where the unit of mass is $\rmn{M_{\sun}}$.
As the disc mass is correlated to the stellar mass according to Equation (\ref{eq:m_d}), Table \ref{tab:M_s2} shows the disc masses and radii that are applied in the simulations in order to faciliate the description of the simulations. In the following, only the mass of the perturbing star is mentioned for each simulation. The corresponding disc masses and radii can be found in Table \ref{tab:M_s2} if not otherwise stated.

\begin{table}
 \begin{center}
\begin{tabular}{|c|c|c|c|c|c|}
\hline
$M_{s2} [\rmn{M_{\sun}}]$ & $M_{d2}[\rmn{M_{\sun}}]$ & $R_{d2}[\rmn{au}]$ \\
\hline
0.5 & 0.004 & 30 \\
1 &  0.01 &  40   \\
5 &  0.11 &   50  \\  
10 & 0.32 &   100 \\
%20 & 0.89 & 200 \\
\hline
\end{tabular}
\end{center}
\caption{Disc masses and radii as function of the perturbing stellar mass.}
\label{tab:M_s2}
\end{table}

%In addition to self-gravity, energy transfer and mass infall onto the disc are neglected.

The  number of gas particles for most of  our simulations was taken to be  $2 \times M_d[M_{\sun}] \times 10^6$. 
 % for more than $150$ orbits in our dimensionless units.
%This length of time was required to reach the time period where the disc parameters do not change significantly anymore, thus have relaxed to smooth evolution again after the stellar encounter. 
In order to run a large sample of simulations, each simulation was stopped 100 orbits / 1120 yr after the pericenter passage of the secondary star-disc system. At that time, the influence of the secondary system is significantly small such that the primary disc inclination, which is our main focus, does not change anymore.  Because of the large computing times, we surrender the study of the long-term evolution of the discs.  

% as we will discuss in Section \ref{sec:longterm}.

\section{Parameter space for realistic encounters} \label{sec:orbit}

In this paper, we study parabolic orbits ($e=1$) for the stellar encounters. Because of the huge parameter space,  we firstly study secondary discs with midplanes being coplanar to the parabolic orbit at simulation start.
The relative orbit of the secondary star to the primary star is a parabolic orbit which can be rotated in space. 
We here apply the same nomenclature as \cite{Xia2016} by defining the pericenter of the orbit in the  $(x,y)$ plane. 
Two two types of rotation are determined by the angles $\beta_p$ (anti-clockwise rotation about the $x$ axis) and $\gamma_p$ (clockwise rotation about the $y$ axis). \citep[See][for more details of the orbital configuration]{Xia2016}. 

As previous analytical and numerical studies have shown \citep[e.g.][]{Ost1994, Lar1997, Thi2005, Xia2016},  the simulation results are expected to strongly depend on the periastron distance between the two star-disc systems. Since the gravitational force decreases with increasing distance according to $r^{-2}$, the mass transfer and the gravitational perturbations between the two discs are expected to decrease rapidly. In order to simulate a close encounter with significant mass transfer between the two discs, the periastron distance should be chosen to be relatively small. 

Every simulation is characterized by the stellar mass ratio $P_m=M_{s2}/M_{s1}$, the disc mass ratio $P_d=M_{d2}/M_{d1}$, the disc radii $R_{d1}$ and $R_{d2}$, the periastron distance $r_p$ and the two angles $\beta_p$ and $\gamma_p$. The orbit of the perturber is computed such that the perturber passes its pericenter at $t=100$ internal time units. Parameters in the following ranges are studied:
$P_m=$[0.5, 1, 5, 10], $R_{d1}=40~\rmn{au}$, $R_{d2}=[30:100]~\rmn{au}$, $r_p=[70: 140]\ \rmn{au}$, $\beta_p$=[0, 45, 90, 135, 180]$\degr$, $\gamma_p$=[0, 45, 90, 135, 180]$\degr$.

\citet{Thi2011} performed simulations with a primary disc of radius $R_{d1}<100 \ \rmn{au}$, a second disc of radius $R_{d2} \approx 400\ \rmn{au}$ and a periapsis of $\approx  500\ \rmn{au}$. In order to set up a similar scenario, we first study close encounters with a periastron distance of $R_{d1}+R_{d2}$. In \citet{Thi2011}, the second disc is more massive with $M_{d2}=0.5\ \rmn{M_{\sun}}$ with a less massive star of $M_{s2}=0.75\ \rmn{M}_{\sun}$ while the primary star is of $1\ \rmn{M_{\sun}}$ and the primary disc mass has $<0.1\ \rmn{M_{\sun}}$. In our simulations, the secondary disc mass is correlated to the second stellar mass according to Equation \ref{eq:m_d}.

\citet{Pic2014} performed simulations of two equal-mass stars both surrounded by equal-mass circumstellar discs undergoing a close encounter. In contrast to their simulations, our survey includes different stellar and disc mass ratios in order to reproduce a more realistic environment in a stellar cluster. The present work can therefore provide a more complete picture of the mass accretion scenario.

\section{Simulation results}\label{sec:results}

\subsection{Single protoplanetary disc}

\begin{figure}
\centering
\includegraphics[width=8cm]{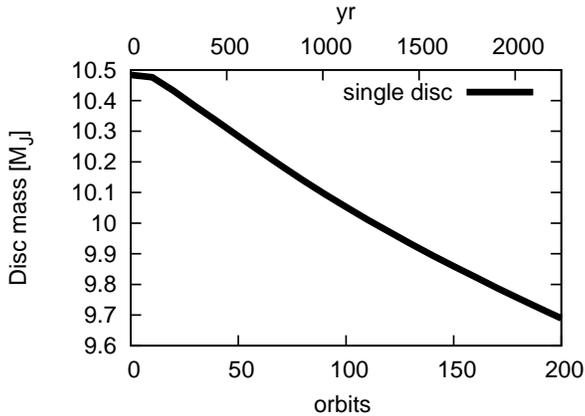}
\caption{The evolution of the disc mass over 200 orbits in the isolated case. The disc mass decreases because of the inward motion of the viscous gas material and subsequent accretion by the central star.}
\label{fig:singledisc}
\end{figure}

In Fig. \ref{fig:singledisc}, the time-dependent evolution of the disc mass in the absence of any perturbations is shown. The hydrodynamical evolution of the viscous gas disc leads to angular momentum loss and hence inward motion of the gas material. In the close vicinity of the central star, the gas material is then accreted. For our chosen parameters, the disc mass decreases from $10.5 \ \rmn{M_J}$ at the simulation start to roughly $9.7\ \rmn{M_J}$ after 200 orbits / 2240 yr. 
The evolution of the single disc mass is relevant for the later analysis and understanding of the disc mass evolution with mass transfer being involved.

\subsection{$\gamma_p=0\degr$, survey of $\beta_p$}

\begin{figure}
\centering
\includegraphics[width=8cm]{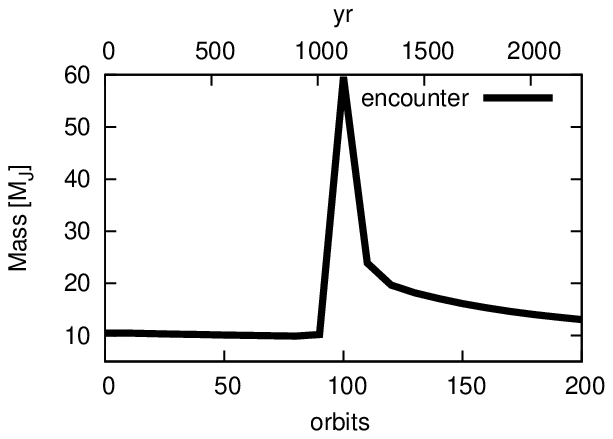} % d3_binary
\includegraphics[width=8cm]{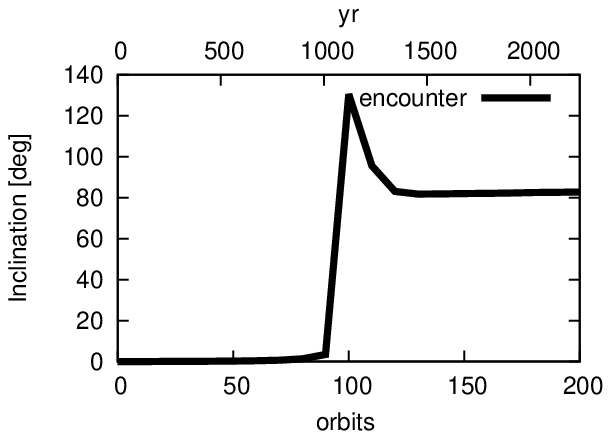} % d3_binary
\caption{Total disc mass (upper panel) and orbital inclination  $i_D$ with respect to the $(x,y)$ plane inside 20 length units as function of time for $P_m=10$, $\beta_p=135\degr$ and $\gamma_p=0\degr$.}
\label{fig:encounter}
\end{figure}

Figure \ref{fig:encounter} shows the total disc mass (upper panel) and inclination  $i_D$ with respect to the $(x,y)$ plane  inside 20 length units  as a function of time for $P_m=10$, $\beta_p=135\degr$ and $\gamma_p=0\degr$.

The disc inclination $i_D$  is obtained from
\begin{eqnarray}
\hspace{-1mm}&i_{D}&= \arccos\left( \frac{ J_{D,z} }{|{\bf J}_{D}|} \right)\ , 
\end{eqnarray}
where $J_{D,z}$ is the $z$-component of the primary disc angular momentum vector ${\bf J}_D$ which is defined as 
\begin{eqnarray}
|{\bf J}_D|=\left| \sum_i m_i ({\bf r}_i \times {\bf v}_i) \right|  \ . \label{eq:J}
\end{eqnarray}
The summation in (\ref{eq:J}) is  taken over all gas particles which are located inside 20 length units and bound to the primary star.

\begin{figure}
\centering
\includegraphics[width=8cm]{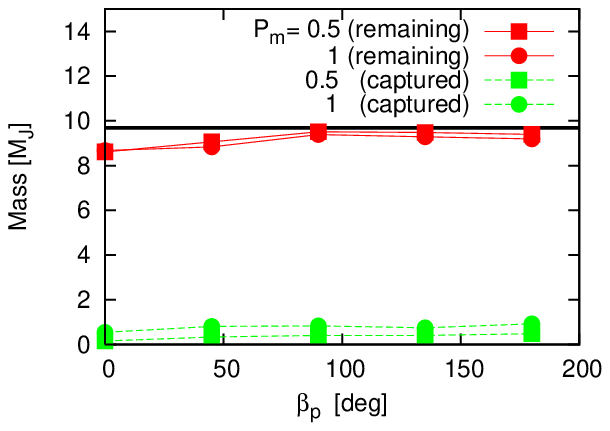}
\includegraphics[width=8cm]{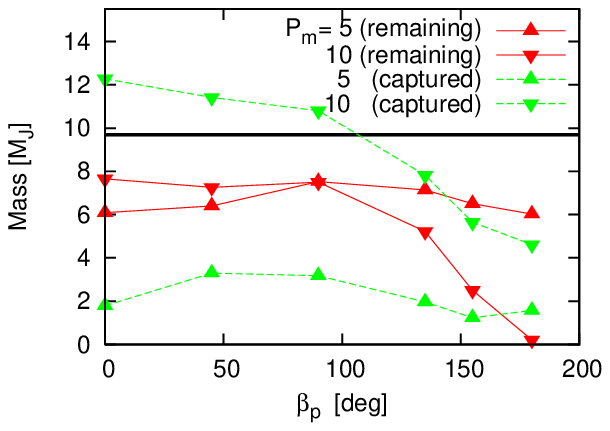}
\caption{Total remaining and captured gas mass inside 20 length units as function of the angle $\beta_p$ for $P_m=0.5$, 1 (upper panel) and 5, 10 (lower panel). For all simulations $\gamma_p=0\degr$. The black horizontal lines indicate the remaining mass of an isolated disc after 200 orbits / 2240 yr.}
\label{fig:beta_m}
\end{figure}

In both plots, the peak at 100 orbits (=1120 yr) marks the pericenter passage of the perturber. 
 Shortly after the pericenter passage, the primary disc virialises and the disc inclination evolves into an almost constant value. The disc mass, in contrast, continues to decrease because of gas accretion onto the central star. This can be verified by analyzing the evolution of the disc and stellar mass. During the period $t=150-200$ orbits in Figure \ref{fig:encounter}, the disc mass decreases by 3 $\rmn{M_J}$. At the same time, the central stellar mass increases by 3 $\rmn{M_J}$. Although the disc mass is evolving continuosly, the disc inclination which is of more interest for our study has reached a constant value so that the simulation can be stopped. In order to simplify the visualization of our parameter surveys, we will hence only show the final values of the disc mass and disc inclination at $t=200$ orbits (=2240 yr) in the following figures.

\begin{figure}
\centering
\includegraphics[width=8cm]{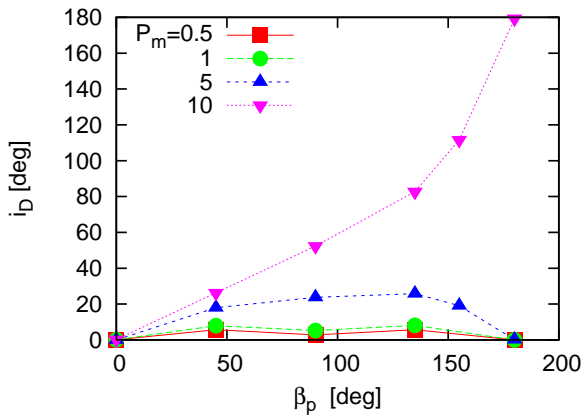}
\caption{Disc inclination $i_D$ with respect to the $(x,y)$ plane of the primary disc inside 20 length units as function of the angle $\beta_p$ for stellar mass ratios in the range [0.5:10] as indicated by the different colors. For all simulations $\gamma_p=0\degr$. }
\label{fig:beta_i}
\end{figure}

Figure \ref{fig:beta_m} shows the total remaining and captured gas mass inside 20 length units for $\gamma_p=0\degr$, different values of $\beta_p$ and stellar mass ratios in the range [0.5:10]. The periastron distance $r_p$ is chosen to be $R_{d1}+R_{d2}$.
The correponding final disc inclination for different values of $\beta_p$ and $P_m$ at $t=200$ is shown in Figure \ref{fig:beta_i}.

For $P_m \leq 1$ / $P_d \leq 1$ (upper panel of Figure \ref{fig:beta_m}), there is little mass transfer between both discs. The primary disc loses a small amount of mass ($\lesssim 1.5\ \rmn{M_J}$) to the secondary disc and accretes less than $1\ \rmn{M_J}$. In all cases, the primary disc inclination does not change significantly and is below $10\degr$. The case $P_m=1$ / $P_d=1$ is comparable to the simulation setup of \cite{Pic2014} who studied two equal-mass stars and discs undergoing a close encounter. Our result is in accordance with the result of \cite{Pic2014} who found final disc inclinations below $10\degr$ in all cases for the particular setup chosen. 

 For $P_m=5$ / $P_d=11$ (lower panel of Figure \ref{fig:beta_m}), mass accretion can increase the primary disc mass by $\lesssim 1 ~ \rmn{M_J}$. For coplanar prograde ($\beta_p=0\degr$) and retrograde ($\beta_p=180\degr$) close encounters, the primary disc loses a significant amount of mass of $\approx  4\ \rmn{M_J}$ to the secondary disc while the accreted mass is only $2\ \rmn{M_J}$. %This can be explained by the fact that in these two extreme cases, both discs are located in one plane and the interaction between the two discs is strongest. 
For $\beta_p=90\degr$, the remaining mass of the primary disc is the largest with $>7\ \rmn{M_J}$. Here, the secondary disc is perpendicular to the primary disc. The primary disc accretes more mass for prograde encounters than for retrograde encounters. However, the final disc inclination is higher for retrograde encounters. For retrograde inclined encounters, the total disc inclination can be increased to $\lesssim 30\degr$. 
 This is because the gas particles which are accreted in the retrograde cases, have a much higher orbital inclination than in the prograde cases. Thus, it is not only the amount of accreted mass which determines the final disc inclination. The orbital inclination of the accreted material also plays a major role for the final disc orientation.

 For $P_m=10$ / $P_d=32$ (lower panel of Figure \ref{fig:beta_m}) , mass accretion is dominant with the accreted mass being higher than the remaining bound disc material for all cases of $\beta_p$.
The prograde cases show less mass loss of the primary disc to the secondary disc and more accretion from the second disc which results in significantly high final disc masses up to $20\ \rmn{M_J}$. For $\beta_p \leq 90\degr$, the final disc inclination is $< 60\degr$.
The retrograde cases have a more destructive nature. The secondary disc accretes a large fraction of the primary disc mass while leaving a moderate mass amount within 20 length units. For $\beta_p=135\degr$, the final disc mass within 20 length units is approximately $13\ \rmn{M_J}$. Here, the accreted mass is higher than the mass which is lost by the primary disc. The final disc inclination is $>80\degr$.

For $\beta_p>135\degr$, we have run two simulations with $\beta_p=155\degr$ and $\beta_p=180\degr$. In both cases, the primary disc loses more material than it has accreted from the secondary disc. At $t=200$ orbits (=2240 yr), the accreted mass is higher than the remaining mass. The accreted mass with a  significantly different angular momentum vector leads to very high final disc inclinations with $i_D>110\degr$. 
We should point out that the case $\beta_p=180\degr$ is very destructive with the primary disc being completely replaced by material from the secondary disc. As our resolution for this case is low, further investigation should be undertaken for a better understanding of a coplanar retrograde encounter. However, due to the large disc mass ratio of $P_d=32$, it is expected that the secondary disc dominates the evolution of both discs after the encounter.

\subsection{$\beta_p=0\degr$, survey of $\gamma_p$}

\begin{figure}
\centering
\includegraphics[width=8cm]{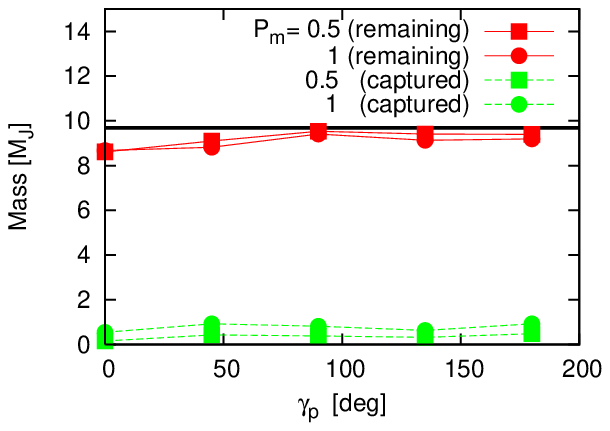}
\includegraphics[width=8cm]{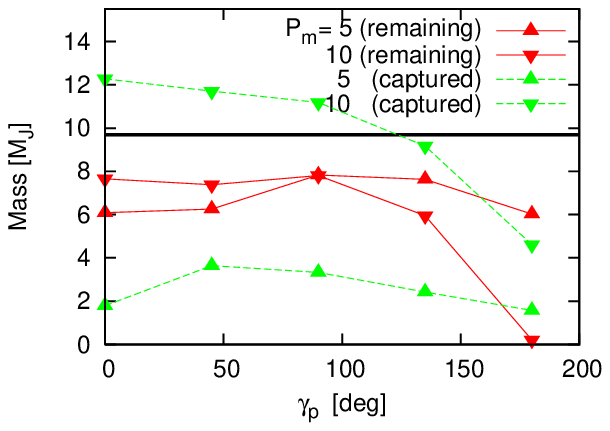}
\caption{ Total remaining and captured gas mass inside 20 length units as function of the angle $\gamma_p$ for $P_m=0.5$, 1 (upper panel) and 5, 10 (lower panel). For all simulations $\beta_p=0\degr$. The black horizontal lines indicate the remaining mass of an isolated disc after 200 orbits / 2240 yr. }
\label{fig:gamma_m}
\end{figure}

 Figure \ref{fig:gamma_m} shows the total remaining and captured gas mass inside 20 length units for $\beta_p=0\degr$, different values of $\gamma_p$ and stellar mass ratios in the range [0.5:10]. The periastron distance $r_p$ is again chosen to be $R_{d1}+R_{d2}$.

\begin{figure}
\centering
\includegraphics[width=8cm]{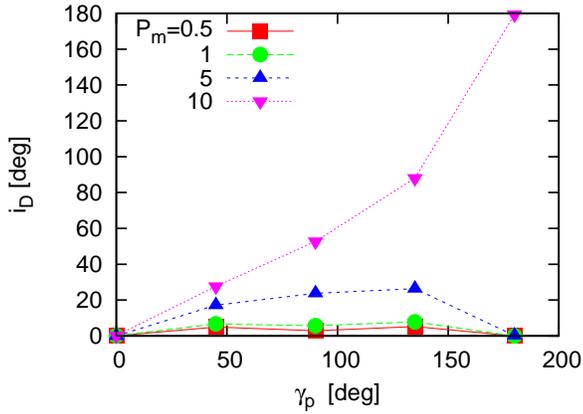}
\caption{Disc inclination $i_D$ with respect to the $(x,y)$ plane of the primary disc inside 20 length units as function of the angle $\gamma_p$ for stellar mass ratios in the range [0.5:10] as indicated by the different colors. For all simulations $\beta_p=0\degr$. }
\label{fig:gamma_i}
\end{figure}

Figure \ref{fig:gamma_i} shows the correponding final disc inclination for different values of $\beta_p$ and $P_m$ at $t=200$ orbits / 2240 yr. 
The results are very similar to the simulations with $\gamma_p=0\degr$ and variations of $\beta_p$.  

 For $P_m \leq 1$ / $P_d \leq 1$ (upper panel of Figure \ref{fig:gamma_m}) , the interaction between the two discs is very weak which leads to little mass transfer between the discs. The accreted mass onto the primary disc is below $1\ \rmn{M_J}$ and the primary disc loses $\lesssim 1.5\ \rmn{M_J}$ in total. 
In all cases, the primary disc inclination again is below $10\degr$ for all studied values of $\gamma_p$. Here, our result is also in accordance with the result of \cite{Pic2014} who found final disc inclinations below $10\degr$ in all cases.

 For $P_m=5$ / $P_d=11$ (lower panel of Figure \ref{fig:gamma_m}), mass accretion can increase the primary disc mass by $\approx  1 ~ \rmn{M_J}$. For $\beta_p=90\degr$, the remaining mass of the primary disc is the largest with $\approx  8\ \rmn{M_J}$. The primary disc accretes more mass through prograde encounters than through retrograde encounters. Similar to the previous survey of $\beta_p$, the final disc inclination is higher for retrograde encounters because of the higher orbital inclinations of the accreted gas particles. For retrograde inclined encounters, the total disc inclination is $\lesssim 30\degr$. 
 
 For $P_m=10$ / $P_d=32$  (lower panel of Figure \ref{fig:gamma_m}), the results are the same as those for variations of  $\beta_p$. In the prograde cases, one can find less mass loss of the primary disc to the secondary disc and more accretion from the second disc which results in significantly high final disc masses up to $20\ \rmn{M_J}$. For $\gamma_p \leq 90\degr$, the final disc inclination is $< 60\degr$. In the retrograde cases,the secondary disc accretes a large fraction of the primary disc mass while leaving a moderate mass amount within 20 length units. For $\beta_p=135\degr$, the final disc mass within 20 length units is $15\ \rmn{M_J}$. In comparison to the case ($\gamma_p=0\degr,\beta_p=135\degr$), the primary disc loses $1\ \rmn{M_J}$ less and accretes $1\ \rmn{M_J}$ more than the previous case. The final disc inclination is $\approx  90\degr$. The case $\gamma_p=180\degr$ is identical to the case $\beta_p=180\degr$ because both simulations represent a coplanar retrograde orbit.

\begin{figure}
\centering
\includegraphics[width=8cm]{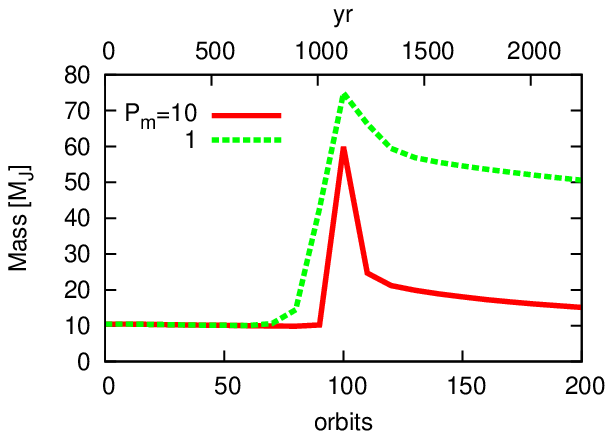}
\includegraphics[width=8cm]{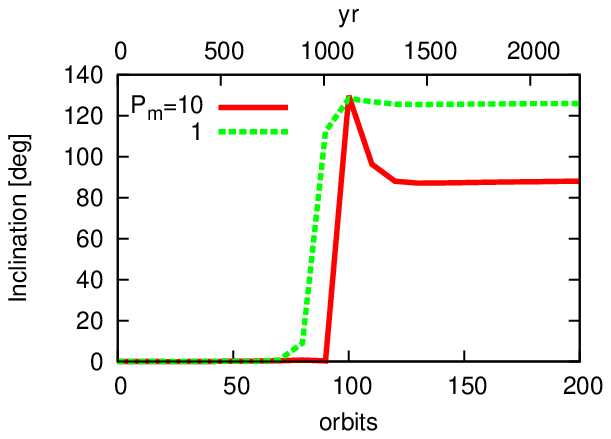}
\caption{Evolution of total gas mass (upper panel) and disc inclination $i_D$ (lower panel) with respect to the $(x,y)$ plane of the primary disc inside 20 length units for $\gamma_p=135\degr$, $\beta_p=0\degr$, $M_{d1}=0.01~ \rmn{M}_{\sun}$, $R_{d1}=40\ \rmn{au}$, $M_{d2}=0.32~ \rmn{M}_{\sun}$, $R_{d2}=100 \ \rmn{au}$, $r_p=140 \ \rmn{au}$. $P_m=10$ (black line) and $P_m=1$ (red line).}
\label{fig:P_m}
\end{figure}

Figure \ref{fig:P_m} shows the evolution of the total gas mass (upper panel) and disc inclination $i_D$ with respect to the $(x,y)$ plane of the primary disc inside 20 length units for $\gamma_p=135\degr$, $\beta_p=0\degr$, $M_{d1}=0.01~ \rmn{M}_{\sun}$, $R_{d1}=40\ \rmn{au}$, $M_{d2}=0.32~ \rmn{M}_{\sun}$, $R_{d2}=100$ au, $r_p=140~ \rmn{au}$ and two different stellar mass ratios $P_m=10$ (red line) and $P_m=1$ (green line). 
In very young and compact stellar clusters \citep[][]{Mar2012}, two protoplanetary discs can undergo a close encounter at an early stage of star formation during which the stellar mass still has not reached its final value. This scenario has been studied by \cite{Thi2011} who assumed a more massive secondary disc with a lower-mass secondary star.
In order to investigate these cases of reduced stellar masses, we have simulated a close encounter between a primary disc of $M_{d1}=0.01~ \rmn{M}_{\sun}$, $R_{d1}=40~ \rmn{au}$ with a solar-type star and a second disc of $M_{d2}=0.32~ \rmn{M}_{\sun}$, $R_{d2}=100~ \rmn{au}$ with a star of only $1\ \rmn{M}_{\sun}$.
It is clearly visible that in the absence of a massive secondary star, the primary disc is capable to accrete a larger amount of gas material from the secondary disc (upper panel) which leads to a larger change of the disc inclination in the primary disc (lower panel). More clearly, in the lower stellar mass case, the final disc mass is $50\ \rmn{M_J}$ while in the case of $P_m=10$, it is only $15\ \rmn{M_J}$. The final disc inclination for $P_m=1$ is $125\degr$ while it is only $<90\degr$ for $P_m=10$. This result indicates that the disc inclination generation is strongly correlated to the amount of accreted mass and the angular momentum of the accreted material.

For close encounters at early stages when the secondary star still has not reached its final mass, we can thus conclude that the mass transfer process is more dominant due to a minor gravitational attraction force of the host star. This can lead to significantly larger final disc inclinations. So far, we have been able to generate highly inclined final discs which can even be retrograde.

Taking into account the result by \cite{Xia2016} who was able to generate disc inclinations $\leq 60\degr$ through single stellar flybys without mass accretion, we can conclude that highly inclined discs can only be generated when we include both the gravitational influence of a passing star and the mass transfer between the two circumstellar discs.

\begin{figure}
\centering
\includegraphics[width=8cm]{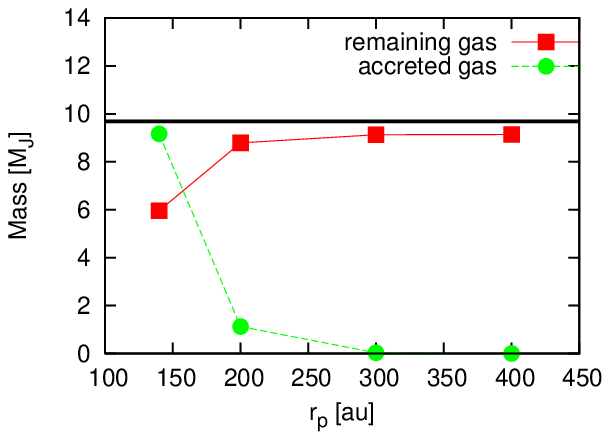}
\includegraphics[width=8cm]{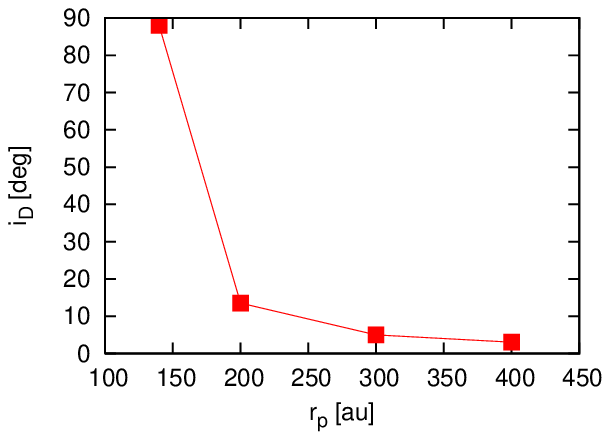}
\caption{Remaining and accreted gas mass (upper panel) and final disc inclination $i_D$ (lower panel) with respect to the $(x,y)$ plane of the primary disc inside 20 length units for $\gamma_p=135\degr$, $\beta_p=0\degr$, $M_{d1}=0.01~ \rmn{M}_{\sun}$, $R_{d1}=40\ \rmn{au}$, $M_{d2}=0.32~ \rmn{M}_{\sun}$, $R_{d2}=100\ \rmn{au}$, $P_m=10$ and different values of $r_p$. The black horizontal line in the upper panel indicates the remaining mass of an isolated disc after 200 orbits / 2240 yr.}
\label{fig:r_p}
\end{figure}

Figure \ref{fig:r_p} shows the remaining (red line) and accreted (green line) gas mass (upper panel) and final disc inclination $i_D$ (lower panel) with respect to the $(x,y)$ plane of the primary disc inside 20 length units for $\gamma_p=135\degr$, $\beta_p=0\degr$, $M_{d1}=0.01~ \rmn{M}\_{\sun}$, $R_{d1}=40\ \rmn{au}$, $M_{d2}=0.32~ \rmn{M}_{\sun}$, $R_{d2}=100\ \rmn{au}$, $P_m=10$ and different values of $r_p$. As shown clearly, mass accretion decreases with increasing $r_p$ which leads to a lower final disc inclination in the primary disc. For $r_p=140\ \rmn{au}$, the total accreted gas mass is $\approx  9\ \rmn{M_J}$ and the disc inclination is almost $90\degr$. For $r_p=200\ \rmn{au}$, the accreted mass is only $\approx  1\ \rmn{M_J}$ and the disc inclination is $\approx  15\degr$. In the  two cases with even higher $r_p$, the final disc inclination is only below $10\degr$. 
While in the two lower cases of $r_p$, mass accretion is the dominant process for the generation of disc inclination, the final disc inclinations of $\leq 10\degr$ in the two higher cases of $i_p$ are a result of the gravitational interaction between the two star-disc systems.
Hence, in order to allow disc inclination generation through mass accretion, the periastron distance of a close encounter is limited to below $r_p\leq 1.5 (R_{d1}+R_{d2})$ for the case $P_m=10$.

\subsection{Disc orientation relative to orbital plane}

\begin{figure}
\centering
\includegraphics[width=8cm]{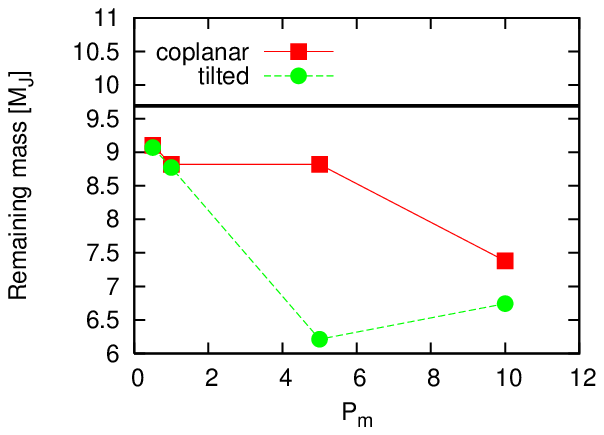}
\includegraphics[width=8cm]{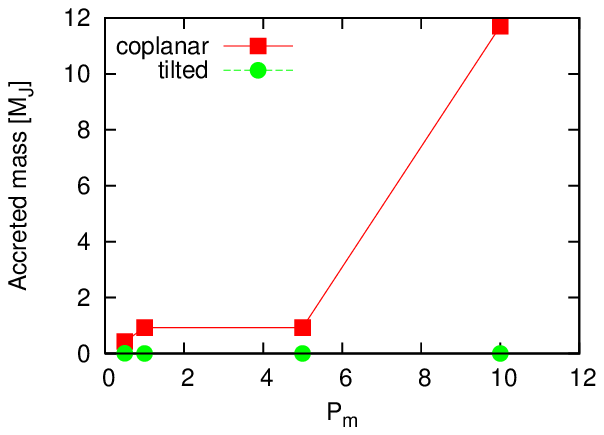}
\caption{Remaining mass (upper panel) and accreted mass (lower panel) inside 20 length units for $\gamma_p=45\degr$, $\beta_p=0\degr$, $M_{d1}=0.01~\rmn{M}_{\sun}$, $R_{d1}=40\ \rmn{au}$ and different values of $P_m$. Red linepoints indicate simulations with the secondary disc being coplanar to the orbital plane while green linepoints indicate simulations with the secondary disc being tilted by $90\degr$ to the orbital plane. The black horizontal line in the upper panel indicates the remaining mass of an isolated disc after 200 orbits / 2240 yr.}
\label{fig:tilt_m}
\end{figure}

\begin{figure}
\centering
\includegraphics[width=8cm]{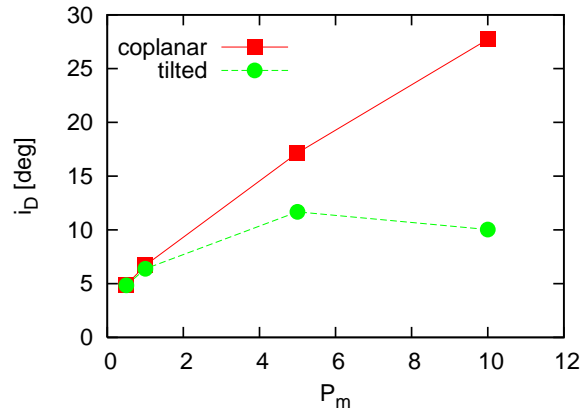}
\caption{Final disc inclination $i_D$ to the $(x,y)$ plane for the same parameters as shown in Fig. \ref{fig:tilt_m}. }
\label{fig:tilt_i}
\end{figure}

In stellar clusters, the three-dimensional orientation of a protoplanetary disc is probably independent of the orbital motion of its host star. With respect to a close encounter, there is probably no correlation between the secondary disc orientation and the orbital plane of the secondary star. A correlation  between stellar spins in two old open clusters has however recently been documented by \citet{Cor2017}. So far, we have assumed that the secondary disc lies in the orbital plane of its host star. Nevertheless, in order to include possible non-vanishing angles between the disc and the orbital plane, a huge parameter space would have to be studied. As the main goal of this paper is to demonstrate the capability of the mass accretion scenario to explain the formation of misaligned Hot Jupiters, we will not perform a full parameter study in this case. Instead, we only show a comparison between simulations with a secondary disc being coplanar to the orbital plane of the second star (red linepoints) and simulations with a tilted second disc by $90\degr$ to the orbital plane in Figure \ref{fig:tilt_m} and \ref{fig:tilt_i}. The specific simulation configurations are $\gamma_p=45\degr$, $\beta_p=0\degr$, $M_{d1}=0.01~\rmn{M}_{\sun}$, $R_{d1}=40\ \rmn{au}$ and different values of $P_m$ .

The coplanar case allows a significant amount of gas being accreted by the primary disc which is in evidence in the highest case of $P_m=10$. There is also mass being lost through the close encounter. In total, the accretion of additional gas with a significantly different angular momentum leads to a considerable change of the total disc inclination up to $<30\degr$. In contrast to the coplanar case, the primary disc reveals a significantly lower remaining mass when the secondary disc is orientated perpendicular to its orbital plane. In the tilted case, the primary disc is not able to accrete mass from the secondary disc. 
In total, the final primary disc mass is mainly determined by the loss of material rather than mass accretion. The lack of mass accretion leads to the conclusion that the final disc inclination up to only $\approx  12\degr$ in Figure \ref{fig:tilt_i} for the tilted case is a result of the purely gravitational interaction between the two systems. 
A more detailed study of possible influences of the disc orientation on the generation of disc inclination is required for a better understanding of the mass accretion scenario and possible impacts of the disc orientation on final disc inclinations. 
As an interesting foretaste, our single example with a disc being perpendicular to the orbital plane of the secondary star has shown that different orientations of the secondary disc can lead to completely different outcomes of the final disc inclinations.
Hence, the disc orientation is expected to be a crucial factor when studying the mass accretion process and the evolution of the disc inclination.

\section{Conclusions} \label{sec:conclusions}

The aim of this work was to test whether the mass accretion scenario is capable to generate significant disc inclinations which would allow the natural formation of misaligned planets and to identify the parameter sets associated with high final disc inclinations.
For this, we have performed  SPH simulations in order to study the influence of mass accretion on the evolution of viscous protoplanetary discs. 
We have run close encounter simulations with two stars each surrounded by their own circumstellar disc. In order to determine possible parameter ranges for significant disc inclination generation, we have run a first systematic parameter survey. 
As the parameter space is huge for the present problem, we have chosen the following characteristics and parameter ranges for the first survey. We have assumed in most simulations that the second disc mid-plane coincides with the plane of the parabolic orbit. The stellar mass ratio $P_m=M_ {s2}/M_{s1}$ is in the range $[0.5:10]$, the disc masses $M_{d1}$ and $M_{d2}$ are in the range $[0.004:0.32]~ M_{\sun}$. In accordance to the recent observational data, we have applied the disc-star mass relation of $M_d=0.01 M_*^{1.5}$ (Eq. \ref{eq:m_d}). The disc radii are in the range $[30:100]~\rmn{au}$. The periastron distance is $r_p=[70:140]~\rmn{au}$. The two angles $\beta_p$ and $\gamma_p$ characterize the three-dimensional orientation for the orbit of the secondary star and disc and are studied in the full range. 
Although the choice of the relatively small disc radii in comparison to observed disc radii of up to several 100 au and the resulting small periastron distances are mainly due to numerical resolutional reasons, future simulations with higher resolutions and larger disc radii are expected to even favour the studied accretion scenario because the cross section of the close encounters increases with increasing disc radii.

%We studied the two extreme cases of $(\beta_p=0\degr$, $\gamma_p\neq 0\degr)$ and $(\beta_p\neq 0\degr, \gamma_p=0\degr)$. Both cases showed the same outcome with respect to the disc evolution. 

For comparison to the simulations shown by \citet{Pic2014}, we have studied close encounters with $P_m = 1$ / $P_d = 1$. For this case, the interaction is weak which manifests in little mass transfer between the two discs. The resulting disc inclination is below $10\degr$ which is in accordance to the results by \citet{Pic2014}. 
As in stellar clusters, stars with significantly different masses undergo close encounters during their evolution, $P_m = 1$ / $P_d = 1$ does not represent the whole parameter space and thus, general conclusions cannot be drawn by sole studies of $P_m\leq 1$. 

Hence, in our survey, the cases $P_m=5$ / $P_d=11$ and $P_m=10$ / $P_d=32$ have also been studied. Both cases show a significantly strong interaction during the close encounter with masses up to $12\ \rmn{M_J}$ being accreted.
Although the primary disc succeeds in accreting more mass in the prograde cases, the final disc inclination is higher in the retrograde cases. This is due to the higher angular momentum of the accreted mass than in the prograde cases.  
By applying the mass accretion scenario, we succeeded in generating disc inclinations in the whole range $[0\degr:180\degr]$.

In order to study a close encounter during an early evolutionary stage which has also been studied by \citet{Thi2011}, we have performed a simulation with a massive secondary disc of $M_{d2}=0.32\ \rmn{M_J}$ which surrounds a central star of only $M_{s2}=1\ \rmn{M_{\sun}}$. This run is characterized by a stellar mass ratio of only $P_m=1$ and a disc mass ratio of $P_d=32$. The presence of a less massive secondary star allows the primary disc to accrete a significantly larger amount of mass from the secondary disc which leads to a final disc inclination of $\approx  125\degr$. This is by $30\degr$ more than in the case of a massive secondary star with $P_m=10$. 
In summary, the final disc inclinations of the parameter survey with the relation $M_d=0.01 M_*^{1.5}$ are not absolute lower limits of inclinations because final secondary stellar masses were assumed in these simulations. In the cases with a more massive secondary disc, lower secondary proto-stellar masses favour an even higher final disc inclination. 
The key factor for the generation of significant disc inclination is connected to the condition that sufficient gas material with different angular momentum is accreted. 

In order to study the possible influence of the periastron distance $r_p$ on the simulations, we have performed runs with $r_p=[140:400]\ \rmn{au}$. For $r_p<1.5 (R_{d1}+R_{d2})$, the accreted mass is $\geq 1\ \rmn{M_J}$ and the final disc inclination is above $10\degr$. For larger values of $r_p$, no additional mass is accreted onto the primary disc and the final disc inclination with $<10\degr$ is a result of the gravitational influence of the secondary star-disc system.

To conclude, our simulation results indicate that the mass accretion scenario is capable to generate disc inclinations in the whole range. Thereby, highly inclined discs can be generated for high disc mass ratios and $r_p< 1.5 (R_{d1}+R_{d2})$ as well as for encounters with a less evolved secondary star plus disc. By implication, encounters between accreting proto-stars are likely to readily lead to significantly inclined discs through mutual accretion. In comparison to a purely gravitational stellar flyby without mass transfer \citep[][]{Xia2016}, the present model can be seen as an improvement towards a more complete picture. Mass transfer also leads to shrinkage of a previously existing disc or of a young planetary system which naturally leads to misaligned Hot Jupiters \citep[][]{Thi2011}. More distant encounters may lead to the generation of spiral density waves in the discs such that gravitational collapse and/or dust coagulation may be enhanced forming outer planetary-system objects \citep[][]{Thi2005, Thi2010}.  

In contrast to the previous simulations with the secondary disc being coplanar to the orbital plane, we have also run a simulation with the secondary disc being perpendicular to the orbital plane of the second star. In the tilted case, the mass loss of the primary disc is significantly different to the coplanar case. At the same time, the primary disc is not capable to accrete gas from the secondary disc in this configuration. 
Thus, the relative orientation of the two discs as well as the relative orbit of the second star together play an important role during close encounters.

%%%%%%%%%%%% P_m in clusters, occurence rate of misaligned systems %%%%%%%%%%%%%%%
The results provided here are not only restricted to primary stellar masses of  $1\ \rmn{M_{\sun}}$. Moreover, they are dependent on the stellar and disc mass ratios and thus applicable to a large sample of stellar clusters. The probability and frequency of close encounters have been determined by several groups in the recent years \citep[e.g.][]{Vin2015, Thi2005, Vin2016}. In fig. 2 of \citet{Vin2016}, the number of fly-bys per star which results in a reduction of the disc size varies between 1 and 5 as a function of the actual cluster model. Ignoring the condition of reducing the disc size, the absolute number of fly-bys per star is even higher. In addition, from fig.1 in \citet{Thi2005},  it is evident that encounters for which mass transfer between the accretion disks is important (100-500AU) are likely in typical embedded clusters forming throughout the Milky Way. These are a few hundred $\rmn{M_{\sun}}$ heavy and have half-mass radii of about $0.5\,$pc \citep[][]{Mar2012}. For young stellar clusters, \cite{Thi2005} determined the propability for close encounters to be $10-32 \%$. This propability can increase up to $50\%$ for more massive young clusters and ONC-type clusters.  
%A rough estimate of the lifetimes of massive accretion disks (when the central star has less than about 95~\% of its mass \citep[see][]{Wuch2003}), is $0.1\,$Myr. The close-encounter rate then implies that per such embedded cluster about $1\,$\% of stars may accrete gas from a passing neighbour such that it affects the forming planetary system.  Thus roughly $1\,$\% of all Solar-type stars should have a misaligned planetary system.
 So far, the sample of confirmed planets with significant misalignments is comparably small and includes possible observational selection effects.  A detailed numerical assessment of the occurrence of misaligned planetary systems would need a dynamical population synthesis calculation in which stars form in embedded clusters with circumstellar disks with the corresponding lifetimes in order to then add-up all cases where mass transfer occurs. This is beyond the scope of the present study.
Given the current uncertainties of the relative frequency of existing misaligned planetary systems and the lack of comparison between theoretical and observational quantities, we suggest that mass transfer between passing very young accretion disks may be an important contribution to the population of misaligned planetary systems.

%%%%%%%%%%%% P_m in clusters, occurence rate of misaligned systems %%%%%%%%%%%%%%%

As discussed in this paper, the relative orientation of the secondary disc with respect to the orbital plane of its host star is crucial for the generation of inclined discs. This will be studied in more detail in future work.
The mass accretion scenario is generally not restricted to only disc-disc encounters, but could also take place in a stellar cluster during an encounter of a star-disc system with any arbitrary gaseous medium, such as e.g. a Bok globule. As the geometry and velocity distribution of the gaseous medium is different to those of a protoplanetary disc, such an encounter could alter the evolution and orientation of the gaseous disc in a different way from a disc-disc encounter.  This scenario will also be investigated in a future project.

%\begin{acknowledgements} %aa

\section*{Acknowledgments} %mnras
Xiang-Gruess acknowledges support of the University of Bonn through the Annemarie Schimmel fellowship.
Simulations were performed using the Fornax Cluster of the Department of Applied Mathematics and Theoretical Physics (DAMTP) at the University of Cambridge (UK).
%\end{acknowledgements} %aa

%\bibliographystyle{aa}
%\bibliography{mybib} % aa

%\begin{appendix}
  
%\section{Precession angle and resulting tilt of the disc midplane} \label{ap:precession}

%For any arbitrary precession angle, we can derive the general expression of the tilt of the disc midplane.

%\end{appendix}

\label{lastpage}

\end{document}